\documentclass[a4paper]{article}

\usepackage{INTERSPEECH2021}
\usepackage{url}
\usepackage{multirow}
\usepackage{adjustbox}
\usepackage{dblfloatfix}
\usepackage{balance}

\usepackage{hyperref}

\DeclareMathOperator*{\secondmax}{max_{2nd}}

\title{End-to-end speaker segmentation for overlap-aware resegmentation}
\name{Herv\'e Bredin$^1$ \& Antoine Laurent$^2$\thanks{This work was granted access to the HPC resources of IDRIS under the allocation AD011012177 made by GENCI, and was partly funded by the French National Research Agency (ANR) through the PLUMCOT (ANR-16-CE92-0025) and the GEM (ANR-19-CE38-0012) projects.}}
\address{
  $^1$IRIT, Université de Toulouse, CNRS, Toulouse, France\\
  $^2$LIUM , Université du Mans, France}
\email{herve.bredin@irit.fr, antoine.laurent@univ-lemans.fr}

\begin{document}

\maketitle
 
\begin{abstract}
Speaker segmentation consists in partitioning a conversation between one or more speakers into speaker turns. Usually addressed as the late combination of three sub-tasks (voice activity detection, speaker change detection, and overlapped speech detection), we propose to train an end-to-end segmentation model that does it directly.
Inspired by the original end-to-end neural speaker diarization approach (EEND), the task is modeled as a multi-label classification problem using permutation-invariant training. The main difference is that our model operates on short audio chunks (5 seconds) but at a much higher temporal resolution (every 16ms). Experiments on multiple speaker diarization datasets conclude that our model can be used with great success on both voice activity detection and overlapped speech detection. Our proposed model can also be used as a post-processing step, to detect and correctly assign overlapped speech regions. Relative diarization error rate improvement over the best considered baseline (VBx) reaches 17\% on AMI, 13\% on DIHARD~3, and 13\% on VoxConverse.
\end{abstract}

\section{Introduction}

The speech processing community relies on term \textit{segmentation} to describe a multitude of tasks: from classifying the audio signal into three classes \{\textit{speech}, \textit{music}, \textit{other}\}, to detecting breath groups, localizing word boundaries, or even partitioning speech regions into phonetic units. 
On this coarse-to-fine time scale, speaker segmentation lies somewhere between \{\textit{speech}, \textit{non-speech}\} classification and breath groups detection. It consists in partitioning speech regions into smaller chunks containing speech from a single speaker. It has been addressed in the past as the combination of several sub-tasks. First, voice activity detection (VAD) removes any region that does not contain speech. Then, speaker change detection (SCD) partitions remaining speech regions into speaker turns, by looking for time instants where a change of speaker occurs~\cite{Yin2017}. From a distance, this definition of speaker segmentation may appear clear and unambiguous. However, when looking more carefully, lots of complex phenomena happen in real-life spontaneous conversations -- overlapped speech, interruptions, and backchannels being the most prominent ones. Therefore, researchers have started working on the overlapped speech detection (OSD) task as well~\cite{Charlet2013, Kunesova2019, Bullock2020}. \\

\noindent\textbf{End-to-end speaker segmentation.} Instead of addressing voice activity detection, speaker change detection, and overlapped speech detection as three different tasks, our first contribution is to train a unique end-to-end speaker segmentation model whose output encompasses the aforementioned sub-tasks. This model is directly inspired by recent advances in end-to-end speaker diarization and, in particular, the growing \textit{End-to-End Neural Diarization} (EEND) family of approaches developed by \textit{Hitachi}~\cite{Fujita2019Interspeech,eend_attention,eend_multitask}. The proposed segmentation model is better than (or at least on par with) several voice activity detection baselines, and sets a new state of the art for overlapped speech detection on all three considered datasets: AMI Mix-Headset~\cite{ami}, DIHARD 3~\cite{ryant2020, DIHARD3_Evaluation_Plan}, and VoxConverse~\cite{voxconverse}. We did not run speaker change detection experiments. \\

\noindent\textbf{Overlap-aware resegmentation.} Our second contribution relates to the problem of assigning detected overlapped speech regions to the right speakers. While a few attempts have been made in the past~\cite{Bullock2020, landini2021voxconverse}, it remains a very difficult problem for which a simple heuristic baseline has yet to be beaten~\cite{otterson2007efficient}. We show, through extensive experimentation, that our segmentation model consistently beats this heuristic when turned into an overlap-aware resegmentation module -- setting a new state of the art on the AMI dataset when combined with the \textit{VBx} approach. \\

\noindent\textbf{Reproducible research.} Last but not least, our final contribution consists in sharing the pretrained model and integrating it into \textit{pyannote} open-source library for reproducibility purposes: {\footnotesize{\texttt{\href{https://huggingface.co/pyannote/segmentation}{huggingface.co/pyannote/segmentation}}}}. Expected outputs of the proposed approaches (VAD, OSD, and resegmentation) are also available at this address in {\footnotesize{\texttt{RTTM}}} format to facilitate future comparison.

\section{End-to-end speaker segmentation}

\begin{figure}[t]
  \centering
  \includegraphics[width=\linewidth]{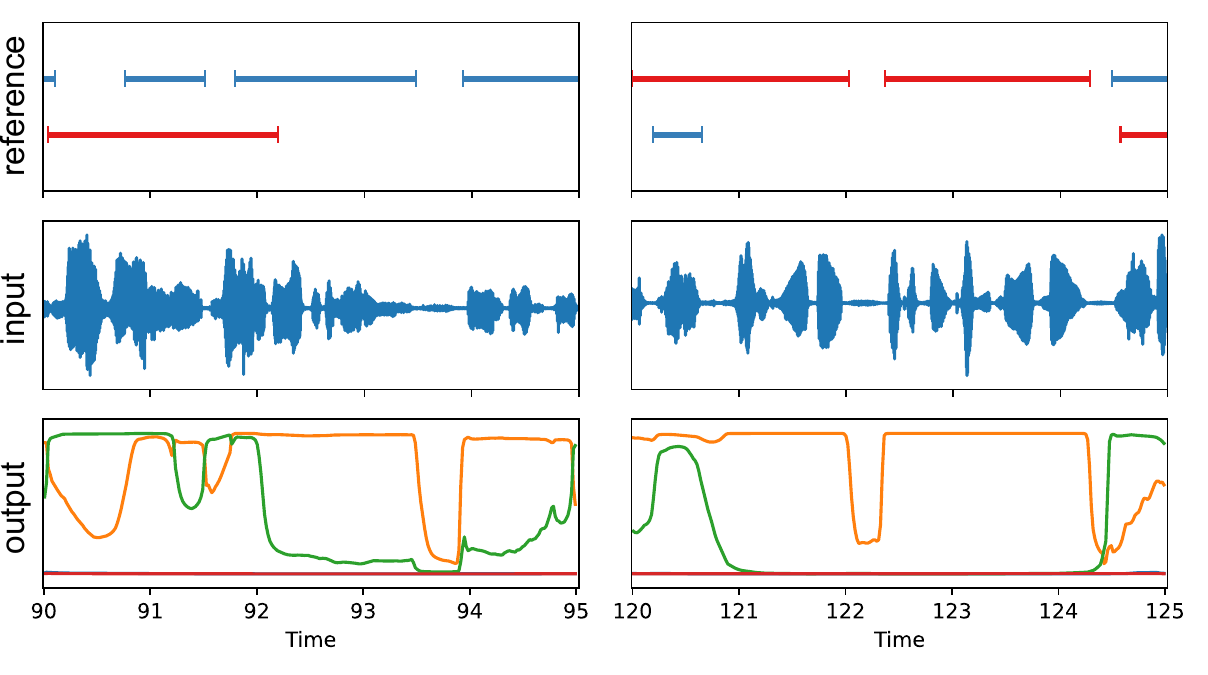}
  \caption{Actual outputs of our model on two 5s excerpts from the same conversation between two speakers (source: file {\footnotesize\texttt{DH\_EVAL\_0035.flac}} in DIHARD3 dataset). Top row shows the reference annotation. Middle row is the audio chunk ingested by the model. Bottom row depicts the raw speaker activations, as returned by the model. Thanks to permutation-invariant training, notice how the blue speaker corresponds to the orange activation on the left and to the green one on the right.}
  \label{fig:two_chunks}
\end{figure}

Like in the original EEND approach~\cite{Fujita2019Interspeech}, the task is modeled as a multi-label classification problem using permutation-invariant training. As depicted in Figure~\ref{fig:two_chunks}, the main difference is that our model operates on short audio chunks (5 seconds) but at a much higher temporal resolution (around every 16ms). Processing short audio chunks also implies that the number of speakers is much smaller and less variable than with the original EEND approach (dealing with whole conversations) -- making the problem easier to address. For instance, we found that $99\%$ of every possible 5s chunks in the training set (later defined in Section~\ref{sec:experiments}) contained less than $K_\text{max} = 4$ speakers.

\subsection{Permutation-invariant training}

Given an audio chunk $\textbf{X}$, its reference segmentation can be encoded into a sequence of $K_\text{max}$-dimensional binary frames $\mathbf{y} = \{ \mathbf{y_1}, \ldots, \mathbf{y_T}\}$ where $\mathbf{y_t} \in \{0, 1\}^{K_\text{max}}$ and $y_t^k= 1$ if speaker $k$ is active at frame $t$ and $y_t^k= 0$ otherwise. We may arbitrarily sort speakers by chronological order of their first activity but any permutation of the $K_\text{max}$ dimensions is a valid representation of the reference segmentation. Therefore, the binary cross entropy loss function $\mathcal{L}_\text{BCE}$ (classically used for such multi-label classification problems) has to be turned into a permutation-invariant loss function $\mathcal{L}$ by running over all possible permutations $\text{perm}(\mathbf{y})$ of $\mathbf{y}$ over its $K_\text{max}$ dimensions:
\begin{equation}
\label{eq:pit}
\mathcal{L}\left(\mathbf{y}, \mathbf{\hat{y}}\right) = \min_{\text{perm}(\mathbf{y})} \mathcal{L}_\text{BCE} \left(\text{perm}(\mathbf{y}), \mathbf{\hat{y}}\right)
\end{equation}
with $\mathbf{\hat{y}} = f(\mathbf{X})$ where $f$ is our segmentation model whose architecture is described later in the paper. In practice, for efficiency, we first compute the $K_\text{max} \times K_\text{max}$ binary cross entropy losses between all pairs of $\mathbf{y}$ and $\mathbf{\hat{y}}$ dimensions, and rely on the Hungarian algorithm to find the permutation that minimizes the overall binary cross entropy loss. 

\subsection{On-the-fly data augmentation}
\label{ssec:augmentation}
For training, 5s audio chunks (and their reference segmentation) are cropped randomly from the training set. To increase variability even more, we rely on on-the-fly random data augmentation. The first type of augmentation is additive background noise with random signal-to-noise ratio. Inspired by our previous work on overlapped speech detection~\cite{Bullock2020}, the second type of augmentation consists in artificially increasing the amount of overlapping speech. To do that, we sum two random 5s audio chunks with random signal-to-signal ratio (and merge their reference segmentation accordingly). Resulting chunks whose number of speakers is higher than $K_\text{max}$ are not used for training.

\subsection{Segmentation} 
\label{ssec:segmentation}

\begin{figure}[t]
  \centering
  \includegraphics[width=\linewidth]{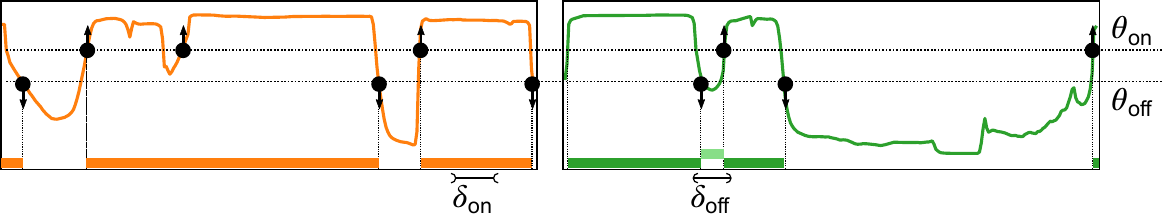}
  \caption{To obtain the final binary segmentation, speaker activations are post-processed with $\theta_{\text{on}} / \theta_{\text{off}}$ hysteresis thresholding, then filling gaps shorter than $\delta_\text{off}$ (light green region in right example) and finally removing active regions shorter than  $\delta_\text{on}$ (does not happen in these examples).}
  \label{fig:thresholds}
\end{figure}

Once trained, the model can be used for segmentation purposes or any sub-tasks by a simple post-processing of its output speaker activations:
\begin{itemize}
    \item for \textbf{segmentation} or \textbf{speaker change detection}, a single $\theta = 0.5$ binarization threshold already gives decent results, but one can get even better performance by using a slightly more advanced post-processing borrowed from~\cite{Gelly2018} and summarized in Figure~\ref{fig:thresholds}.
    \item for \textbf{voice activity detection}, we start by computing the maximum activation over the $K_\text{max}$ speakers:
    \begin{equation}
    \label{eq:vad}
    \hat{y}_t^\text{VAD} = \max_{k} \hat{y}_t^k
    \end{equation}
and, then only, apply the aforementioned post-processing on resulting mono-dimensional $\mathbf{\hat{y}^\text{VAD}}$.
    \item for \textbf{overlapped speech detection}, since at least two speakers need to be active simultaneously to indicate overlapping speech, we compute the second highest (denoted $\secondmax$) activation:
\begin{equation}
\label{eq:osd}
  \hat{y}_t^\text{OSD} = \secondmax_{k} \hat{y}_t^k  
\end{equation}
and post-process the resulting mono-dimensional $\mathbf{\hat{y}^\text{OSD}}$ with the same approach.
\end{itemize}

\begin{table*}[b]
  \caption{Voice activity detection // FA = false alarm rate (\%) / Miss. = missed detection rate (\%)}
  \label{tab:vad}
  \centering
  \begin{adjustbox}{width=0.7\textwidth}
  \begin{tabular}{l|cc|c|cc|c|cc|c}
    \toprule
    \multirow{2}{*}{\textbf{VAD}} & \multicolumn{3}{|c|}{\textbf{AMI}~\cite{ami, vbx}} & \multicolumn{3}{|c|}{\textbf{DIHARD 3}~\cite{ryant2020}} & \multicolumn{3}{|c}{\textbf{VoxConverse}~\cite{voxconverse}} \\ 
    & FA & Miss. & FA+Miss. & FA & Miss. & FA+Miss. & FA & Miss. & FA+Miss. \\
    \midrule
    \textit{silero\_vad} & 9.4 & 1.7 & 11.0 & 17.0 & 4.0 & 21.0 & 3.0 & 1.1 & 4.2 \\
    \textit{dihard3}~\cite{ryant2020} & \tiny{NA} & \tiny{NA} & \tiny{NA} & 4.0 & 4.2 & 8.2 & \tiny{NA} & \tiny{NA} & \tiny{NA} \\
    \textit{Landini et al.}~\cite{landini2021voxconverse} & \tiny{NA} & \tiny{NA} & \tiny{NA} & \tiny{NA} & \tiny{NA} & \tiny{NA} & 1.8 & 1.1 & 3.0 \\
    \textit{pyannote 1.1}~\cite{Bredin2020}  & 6.5 & 1.7 & 8.2 & 4.1 & 3.8 & 7.9 & 4.5 & 0.3 & 4.8 \\
    \midrule
    \textbf{Ours} -- \textit{pyannote 2.0} & 3.6 & 3.2 & \textbf{6.8} & 3.9 & 3.3 & \textbf{7.3} & 1.8 & 0.8 & \textbf{2.5} \\
    \bottomrule
  \end{tabular}
  \end{adjustbox}
\end{table*}

\subsection{Overlap-aware resegmentation}

\begin{figure}[t]
  \centering
  \includegraphics[width=\linewidth]{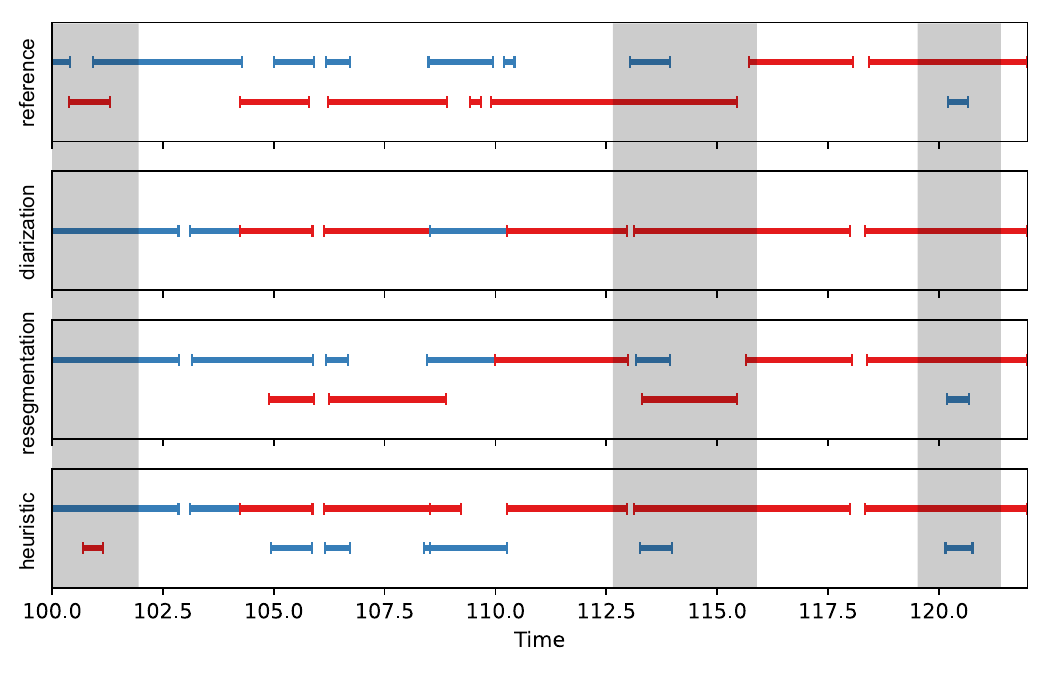}
  \caption{Effect of the proposed overlap-aware resegmentation approach (third row) on the \textit{VBx} diarization baseline (second row). We highlight three regions where the heuristic performs better ($t\approx 100s$), same ($t \approx 120s$), or worse ($t \approx 115$s) than the proposed approach (source: file {\footnotesize\texttt{DH\_EVAL\_0035.flac}} in DIHARD3 dataset).}
  \label{fig:resegmentation}
\end{figure}

While a growing number of diarization approaches do try and take overlapped speech into account~\cite{eend_multitask}, the most dependable ones (like the \textit{VBx} approach~\cite{vbx} used in Figure~\ref{fig:resegmentation}) still assume internally that at most one speaker is active at any time. There is therefore a need for a post-processing step that assigns multiple speaker labels to overlapped speech regions~\cite{Bullock2020,eend_postprocessing}. 

Given an existing speaker diarization output (with $K$ speakers) encoded into a sequence of $K$-dimensional binary frames $y_t^\text{DIA}$, we propose to use the segmentation model as a local, overlap-aware, resegmentation module. The segmentation model is applied on a 5s-long window sliding over the whole file. At each step, we find the permutation of the speaker activations $\mathbf{\hat{y}}$ that minimizes the binary cross entropy loss with respect to $\mathbf{y}^\text{DIA}$. Permutated sliding speaker activations are then aggregated over time and post-processed with the threshold-based approach introduced in Section~\ref{ssec:segmentation}.

\section{Experiments}
\label{sec:experiments}

\noindent\textbf{Datasets and partitions. } We ran experiments and report results on three speaker diarization datasets, covering a wide range of domains:\\ 

\noindent\textit{DIHARD3} corpus~\cite{ryant2020,DIHARD3_Evaluation_Plan} does not provide a \textit{training} set. Therefore, we split its \textit{development} set into two parts: 192 files used as \textit{training} set, and the remaining 62 files used as a smaller \textit{development} set. The latter is simply referred to as \textit{development} sets in the rest of the paper. When defining this split (shared at {\footnotesize{\texttt{\href{https://huggingface.co/pyannote/segmentation}{huggingface.co/pyannote/segmentation}}}}), we made sure that the 11 domains were equally distributed between both subsets. The \textit{evaluation} set is kept unchanged.\\

\noindent\textit{VoxConverse} does not provide a \textit{training} set either~\cite{voxconverse}. Therefore, we also split its \textit{development} set into two parts: first 144 files ({\scriptsize\texttt{abjxc}} to {\scriptsize\texttt{qouur}}, in alphabetical order) constitute the \textit{training} set, leaving the remaining 72 files ({\scriptsize\texttt{qppll}} to {\scriptsize\texttt{zyffh}}) for the actual \textit{development} set.\\

\noindent\textit{AMI} provides an official \{\textit{training}, \textit{development}, \textit{evaluation}\} partition of the Mix-Headset audio files~\cite{ami}. While we kept the \textit{development} and \textit{evaluation} sets unchanged, we only used the first 10 minutes of each file of the \textit{training} set, to end up with an actual \textit{training} set similar in size (22 hours) to that of the DIHARD3 (25 hours) and VoxConverse (15 hours) \textit{training} sets. \\

\noindent\textbf{Experimental protocols. } We trained a unique segmentation model using the \textit{composite} training set (62 hours) made of the concatenation of all three \textit{training} sets. The \textit{composite} development set (24 hours) served as validation and was used to decrease the learning rate on plateau and eventually choose the best model checkpoint. At the end of this process, only one segmentation model  is available (not one model per dataset) and used for all experiments.

However, detection thresholds ($\theta_{\text{on}}$, $\theta_{\text{off}}$, $\delta_{\text{on}}$, and $\delta_{\text{off}}$) were tuned specifically for each dataset using their own \textit{development} set because the manual annotation guides differ from one dataset to another, especially regarding $\delta_{\text{off}}$ which controls whether to bridge small intra-speaker pauses.
\noindent For the same reasons, detection thresholds were optimized specifically for each task addressed in the paper:
\begin{itemize}
    \item voice activity detection thresholds are chosen to minimize the detection error rate (i.e. the sum of the false alarm and missed detection rates), with no forgiveness collar around speech turn boundaries;
    \item overlapped speech detection thresholds are chosen to maximize the detection $F_1$-score, with no forgiveness collar either;
    \item for resegmentation, detection thresholds are chosen to minimize the diarization error rate, without forgiveness collar but with overlapped speech regions. This is consistent with DIHARD3 evaluation plan~\cite{DIHARD3_Evaluation_Plan} and AMI \textit{Full} evaluation setup~\cite{vbx}, but not with VoxConverse challenge rules that uses a 250ms collar~\cite{voxconverse}.
\end{itemize}
All metrics were computed using \textit{pyannote.metrics}~\cite{pyannote.metrics} open source Python library.\\

\noindent\textbf{Implementation details. } Our segmentation model ingests 5s audio chunks with a sampling rate of 16kHz (\textit{i.e.} sequences of 80000 samples). 
The input sequence is passed to \textit{SincNet} convolutional layers using the original configuration~\cite{Ravanelli2018} -- except for the stride of the very first layer which is set to 10 (so that \textit{SincNet} frames are extracted every 16ms). Four bidirectional Long Short-Term Memory (LSTM) recurrent layers (each with 128 units in both forward backward directions, and 50\% dropout for the first three layers) are stacked on top of two additional fully connected layers (each with 128 units and leaky ReLU activation) which also operate at frame-level. A final fully connected classification layer with sigmoid activation outputs $K_\text{max}$-dimensional speaker activations between 0 and 1 every 16ms. Overall, our model contains 1.5 million trainable parameters -- most of which (1.4 million) comes from the recurrent layers.

As introduced in Section~\ref{ssec:augmentation}, 50\% of the training samples are made out of the weighted sum of two chunks, with a signal-to-signal ratio sampled uniformly between 0 and 10dB. We also use additive background noise from the MUSAN dataset~\cite{musan} with a signal-to-noise ratio sampled uniformly between 5 and 15dB. 

We train the model with \textit{Adam} optimizer with default \textit{PyTorch} parameters and mini-batches of size 128. Learning rate is initialized at $10^{-3}$ and reduced by a factor of 2 every time its performance on the development set reaches a plateau. It took around 3 days using 4 V100 GPUs to reach peak performance. While we do share the pretrained model at {\footnotesize{\texttt{\href{https://huggingface.co/pyannote/segmentation}{huggingface.co/pyannote/segmentation}}}} for reproducing the results, the whole training process is also reproducible as everything has been integrated into version 2.0 of \textit{pyannote.audio} open-source library~\cite{Bredin2020}.

\section{Results and discussions}

\begin{table*}[th]
  \caption{Overlapped speech detection // FA = false alarm rate (\%) / Miss. = missed detection rate (\%) / $F_1$ = $F_1$-score (\%)}
  \label{tab:osd}
  \centering
    \begin{adjustbox}{width=1\textwidth}
  \begin{tabular}{l|cc|cc|c|cc|cc|c|cc|cc|c}
    \toprule
    \multirow{2}{*}{\textbf{OSD}} & \multicolumn{5}{|c|}{\textbf{AMI}~\cite{ami, vbx}} & \multicolumn{5}{|c|}{\textbf{DIHARD 3}~\cite{ryant2020}} & \multicolumn{5}{|c}{\textbf{VoxConverse}~\cite{voxconverse}} \\ 
    & FA & Miss. & Precision & Recall & $F_1$ & FA & Miss. & Precision & Recall & $F_1$ & FA & Miss. & Precision & Recall & $F_1$ \\
    \midrule
    \textit{Kunesova et al.}~\cite{Kunesova2019} & \tiny{NA} & \tiny{NA} & 71.5 & 46.1 & 56.0 & \tiny{NA} & \tiny{NA} & \tiny{NA} & \tiny{NA} & \tiny{NA} & \tiny{NA} & \tiny{NA} & \tiny{NA} & \tiny{NA} & \tiny{NA} \\
    \textit{Landini et al.}~\cite{landini2021voxconverse} & \tiny{NA} & \tiny{NA} & \tiny{NA} & \tiny{NA} & \tiny{NA} & \tiny{NA} & \tiny{NA} & \tiny{NA} & \tiny{NA} & \tiny{NA} & 10.4 & 71.8 & 73.0 & 28.2 & 40.7 \\
    \textit{pyannote 1.1}~\cite{Bredin2020,Bullock2020} & 51.1 & 12.1 & 63.2 & 87.9 & 73.5 & 48.2 & 45.2 & 53.2 & 54.8 & 54.0 & 130.4 & 17.7 & 38.7 & 82.3 & 52.6 \\
    \midrule
    \textbf{Ours} -- \textit{pyannote 2.0} & 16.9 & 29.4 & 80.7 & 70.5 & \textbf{75.3} & 46.9 & 37.2 & 57.2 & 62.8 & \textbf{59.9} & 26.3 & 24.5 & 74.2 & 75.5 & \textbf{74.8} \\
    \bottomrule
  \end{tabular}
  \end{adjustbox}
\end{table*}

\begin{table*}[th!]
  \caption{Resegmentation // FA = false alarm / Miss. = missed detection / Conf. = speaker confusion / DER = diarization error rate}
  \label{tab:rsg}
  \centering
  \begin{adjustbox}{width=1\textwidth}
  \begin{tabular}{l|l|ccc|c|ccc|c|ccc|c}
    \toprule
    \multirow{2}{*}{\textbf{Baseline}} & \textbf{Overlap-aware} & \multicolumn{4}{|c|}{\textbf{AMI}~\cite{ami, vbx}} & \multicolumn{4}{|c|}{\textbf{DIHARD 3}~\cite{ryant2020}} & \multicolumn{4}{|c}{\textbf{VoxConverse}~\cite{voxconverse}} \\ 
    & \textbf{resegmentation} & FA & Miss. & Conf. & DER & FA & Miss. & Conf. & DER & FA & Miss. & Conf. & DER \\
    \midrule
    \textit{pyannote 1.1}~\cite{Bredin2020} & \_ & 5.0 & 16.2 & 8.5 & 29.7 & 3.4 & 13.2 & 12.6 & 29.2 & 2.0 & 10.1 & 9.5 & 21.5 \\
    & Heuristic~\cite{otterson2007efficient} w/ \textbf{our OSD} & 6.9 & 7.9 & 10.9 & \textbf{25.7} & 6.3 & 8.9 & 12.8 & 28.1 & 2.8 & 7.3 & 10.1 & 20.3 \\
    & \textbf{Ours} -- \textit{pyannote 2.0}& 4.0 & 13.0 & 9.1 & 26.1 & 5.1 & 9.8 & \textbf{10.3} & 25.2 & 2.4 & 3.1 & 9.8 & \textbf{15.4} \\
        \midrule
    \textit{dihard3}~\cite{ryant2020} & \_ & \tiny{NA} & \tiny{NA} & \tiny{NA} & \tiny{NA} & 3.6 & 13.3 & 8.4 & 25.4 & \tiny{NA} & \tiny{NA} & \tiny{NA} & \tiny{NA} \\
    & Heuristic~\cite{otterson2007efficient} w/ \textbf{our OSD} & \tiny{NA} & \tiny{NA} & \tiny{NA} & \tiny{NA} & 6.8 & 8.7 & 8.8 & 24.3 & \tiny{NA} & \tiny{NA} & \tiny{NA} & \tiny{NA} \\
    & \textbf{Ours} -- \textit{pyannote 2.0} & \tiny{NA} & \tiny{NA} & \tiny{NA} & \tiny{NA} & 4.6 & 10.2 & 7.5 & \textbf{22.2} & \tiny{NA} & \tiny{NA} & \tiny{NA} & \tiny{NA} \\
    \midrule
    \textit{VBx}~\cite{vbx} w/ \textbf{our VAD} & \_ & 3.1 & 17.2 & 3.8 & 24.1 & 3.6 & 12.5 & 6.2 & 22.3 & 1.7 & 5.1 & 1.4 & 8.3 \\
    & Heuristic~\cite{otterson2007efficient} w/ \textbf{our OSD} & 5.1 & 8.7 & 6.1 & \textbf{19.9} & 7.0 & 7.8 & 6.4 & 21.2 & 2.7 & 2.1 & 2.0 & \textbf{6.8} \\
        & \textbf{Ours} -- \textit{pyannote 2.0} & 4.3 & 10.9 & 4.7 & \textbf{19.9} & 4.7 & 9.7 & 4.9 & \textbf{19.3} & 2.7 & 2.6 & 1.8 & 7.1 \\

\midrule
\midrule

    Oracle & \textbf{Ours} -- \textit{pyannote 2.0} & 4.7 & 10.0 & 1.4 & 16.1 & 4.6 & 9.8 & 1.8 & 16.2 & 2.6 & 2.5 & 0.6 & 5.7 \\
    \bottomrule
  \end{tabular}
  \end{adjustbox}
\end{table*}

\noindent\textbf{Voice activity detection. } Table~\ref{tab:vad} compares the performance of the proposed voice activity detection approach with the official \textit{dihard3} baseline~\cite{ryant2020}, \textit{Landini}'s submission to VoxConverse challenge~\cite{landini2021voxconverse}, and \textit{pyannote 1.1} VAD models~\cite{Bredin2020}.
The main conclusion is that, despite it being trained for segmentation, our model is better than other models trained specifically for voice activity detection. Note, however, that one should not draw hasty conclusions regarding the performance of \textit{silero\_vad} model~\cite{silero-vad} as it is an off-the-shelf model which was not trained specifically for these datasets. \\

\noindent\textbf{Overlapped speech detection. } Finding good and reproducible baselines for the overlapped speech detection task proved to be a difficult task. We thank \textit{Kunesova et al.}~\cite{Kunesova2019} and \textit{Landini et al.}~\cite{landini2021voxconverse} for sharing the output of their detection pipelines. Results are reported in Table~\ref{tab:osd} that shows that, like for voice activity detection, our segmentation model can be used successfully for overlapped speech detection, even though it was not initially trained for this particular task. It outperforms \textit{pyannote 1.1} overlapped speech detection which we believe was the previous state of the art~\cite{Bullock2020}. \\

\noindent\textbf{Overlap-aware resegmentation. } While our segmentation model was found to be useful for both voice activity detection and overlapped speech detection, post-processing the output of existing speaker diarization pipelines is where it really shines. Table~\ref{tab:rsg} summarizes the resegmentation experiments performed on top of three of them, ranked from worst to best baseline performance: \textit{pyannote 1.1} pretrained pipelines~\cite{Bredin2020}, \textit{dihard3} official baseline~\cite{ryant2020}, and BUT's \textit{VBx} approach~\cite{vbx}. The (admittedly wrong) criterion used for selecting those baselines was their ease of use and reproducibility. Because results reported in~\cite{vbx} for \textit{VBx} baseline rely on oracle voice activity detection and the shared code base does not provide an official voice activity detection implementation, we used our own (marked as \textbf{Ours} in Table~\ref{tab:vad}) and applied \textit{VBx} on top of it. Our proposed resegmentation approach consistently improves the output of all baselines on all datasets. Relative diarization error rate improvement over the best baseline (\textit{VBx}) reaches 17\% on AMI, 13\% on DIHARD, and 13\% on VoxConverse.

For comparison purposes, we also implemented a heuristic that consists in assigning detected overlapped speech regions to the two nearest speakers in time~\cite{otterson2007efficient}. Despite its simplicity, this heuristic happens to be a strong baseline, very difficult to beat in practice~\cite{landini2021voxconverse}. Yet, our proposed resegmentation approach outperforms it for all but two experimental conditions (for which  the heuristic is better only by a small margin). A closer look at the speaker confusion error rates shows that our approach is significantly better at identifying overlapping speakers. This is confirmed by the low speaker confusion error rates obtained when we apply it on top of an oracle diarization (with $\mathbf{y}^\text{DIA} = \mathbf{y}$): only 1.4\%, 1.8\%, and 0.6\% of speech are re-assigned incorrectly on AMI, DIHARD and VoxConverse respectively. Figure~\ref{fig:resegmentation} provides a qualitative sneak peak at their respective behavior on a short 20 seconds excerpt. In particular, it appears that the two (heuristic and proposed) approaches do behave differently and could complement each other.

\section{Conclusions}

The overall best pipeline reported in this paper is the combination of our voice activity detection, off-the-shelf VBx clustering, and our overlap-aware resegmentation approach, reaching $\text{DER}=19.9\%$ on AMI Mix-Headset using the \textit{full} evaluation setup introduced in~\cite{vbx}, $\text{DER}=19.3\%$ on DIHARD 3 evaluation set (\textit{full} condition, 2.6\% behind the winning submission), and $\text{DER}=7.1\%$ (or $\text{DER}=3.4\%$ with a 250ms forgiveness collar) on VoxConverse \textit{development} set.


Even with a forgiveness collar, missed detection and false alarms are the main source of errors (twice as high as speaker confusion) for all three datasets -- indicating that, despite progress, overlapped speech detection remains an unsolved (and sometimes ill-defined) problem. 

\bibliographystyle{IEEEtran}

\newpage
\balance
\bibliography{reference}

\end{document}